\documentclass[prd,twocolumn, nofootinbib, longbibliography
]{revtex4-1}
\usepackage{tabularx}
\usepackage{float}
\usepackage{datetime}
\usepackage{textpos}
\usepackage{booktabs}
\usepackage{graphicx}

\begin{document}
\title{Evaluating the Gamma-Ray Evidence for Self-Annihilating Dark Matter from the  Virgo Cluster}
\author{Oscar Mac\'\i as-Ram\'\i rez}
\email{oscar.maciasramirez@pg.canterbury.ac.nz}
\author{Chris Gordon}
\author{ Anthony M.\ Brown}
\author{Jenni Adams}
\affiliation{Department of Physics and Astronomy, Rutherford Building, University of Canterbury, Private Bag 4800, Christchurch 8140, New Zealand}

\begin{abstract}
Based on three years of Fermi Large Area Telescope (LAT) gamma-ray data of the Virgo cluster, evidence for an extended emission associated with dark matter pair annihilation in the $b\bar{b}$ channel has been reported by Han et al.~\cite{han}. 
After an in depth spatial and temporal analysis,
we argue that the tentative evidence for a gamma-ray excess from the Virgo cluster is mainly due to the appearance of a population of previously unresolved gamma-ray point sources in the region of interest.  These point sources are not part of the LAT second source catalogue (2FGL), but are found to be above the standard detection significance threshold when three or more years of LAT data is included.
\end{abstract}

\maketitle

\section{Introduction}
\label{sec:introduction}
A plausible dark matter candidate is
a thermal relic arising from a weakly interacting massive particle (WIMP) \cite{cirelli,silk}.  
For supersymmetric extensions of the Standard Model in which \textit{R}-parity is conserved, neutralino particles
 are a natural WIMP candidate. By virtue of their inherent mass range, a few GeV to a few TeV, and allowed gauge couplings they can pair-annihilate generating gamma-ray photons. Currently there is a remarkable international effort to look for such gamma-ray signals in the entire sky~\cite{Aharonian, Abramowski, Acciari} and the Fermi-LAT instrument represents a unique opportunity~\cite{baltz} to fulfill it.

Galaxy clusters are promising astrophysical targets to study the hypothetical annihilation radiation~\cite{pinzke1}. Compared to smaller objects, like dwarf spheroidal galaxies,  clusters possess DM substructures or subhalos which are less affected by tidal stripping, and the uncertainties related with the DM density profiles are usually lower. More importantly, it has also been shown that the presence of subhalos in clusters can considerably enhance the DM luminosities~\cite{pinzke2, Gao}. On the downside, the gamma-ray foreground of clusters may be 
contaminated with point sources such as active galactic nuclei  (AGN) and star-burst~\cite{Acero2, Acciari2}
galaxies. Also, there could be a significant contribution of gamma-rays from cosmic rays (CRs) in clusters, see for example~\cite{pinzke2}.  
Consequently, the task of disentangling a DM signal in the continuum spectrum from the astrophysical noise in galaxy clusters may be difficult~\cite{profumo1, pinzke3}.

Through the use of dedicated cosmological simulations of cluster halos and subhalos from the Phoenix Project~\cite{Gao} and by making some reasonable theoretical assumptions~\cite{pinzke2}, accurate extended dark matter density profiles have been reported~\cite{Gao}. In~\cite{pinzke2} it was found that resolved and unresolved substructures in the inner part of clusters are expected to play a more important role than that of the main smooth Navarro-Frenk-White (NFW) halo. From this it follows that nearby clusters should be the brightest DM radiative sources after the milky way (MW) in the gamma-ray sky. In fact, in Ref.~\cite{pinzke2}, Fornax, M49 and Virgo are found to yield the most intense pair-annihilation radiation as a result of the enhancement provided by their subhalos.         

From the analysis of the first 11 months of publicly available Fermi-LAT data and adopting a smooth NFW density profile, null DM results, from clusters, were obtained~\cite{ackermann1, profumo2}. Instead, constraints on cross sections and masses of WIMPs were derived. It was also the case for the work shown in Ref.~\cite{huang} where, however, an extended DM halo profile (different to the one shown in~\cite{Gao}) and almost 3 years of Fermi data were employed.

\begin{figure}
                \centering
  \includegraphics[width=1\linewidth]{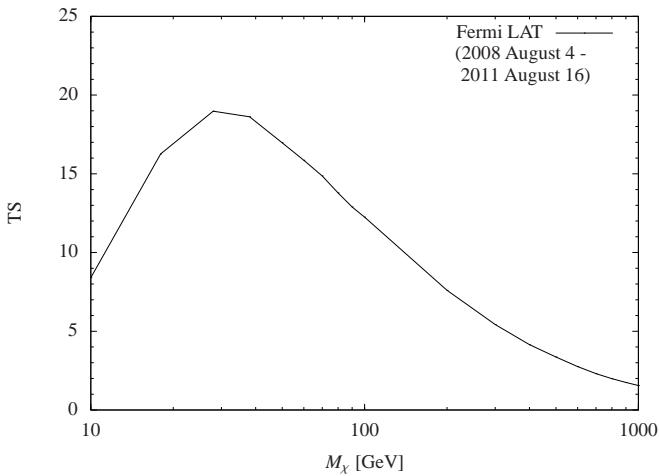}
                	\caption{ \label{fig:hanetalresults} TS values obtained for the Virgo cluster with an extended dark matter density profile plus the 2FGL catalogue point sources and backgrounds. Dark matter annihilation is   assumed to be in the $b\bar{b}$  channel. The CR component is assumed negligible. The fit is based on roughly 3 years of Fermi LAT data. See Ref.~\cite{han} for details.}
\end{figure}

An interesting study of the Virgo cluster by  Han et al.~\cite{han} (HFEGW hereafter) which used a high resolution DM density profile~\cite{Gao} was recently undertaken. They claimed to have found some evidence  for DM annihilation from Virgo  in the $b\bar{b}$ channel for a WIMP mass of about 28 GeV and with a detection significance of 4.4$\sigma$. 
They found that their signal is from a spatially extended region which is also spectrally distinct from the
the radio galaxy (M87) located in the center of the region of interest (ROI). 
They also found that the target region prefers a DM hypothesis over a CR model.
HFEGW used the 2FGL catalogue to determine the list of gamma-ray point sources used in their astrophysical model. We reproduce in Fig.~\ref{fig:hanetalresults} the main results shown in HFEGW for the No-CR model where we make use of their same data set and assumptions.
The test statistic (TS) is defined as in Ref.~\cite{2FGL} 
\begin{eqnarray}\label{tsdef}
TS=2\left[\log \mathcal{L} (\mbox{new source})-\log \mathcal{L} (\mbox{NO-new source})\right]\mbox{,} \qquad 
\end{eqnarray}
where $\mathcal{L}$ stands for the maximum of the likelihood of the data given the model with or without the new source at a certain location of the ROI.
In the large sample limit, under the no source hypothesis, TS has a $\chi^2/2$ distribution with the number of degrees of freedom equal to the number of 
parameters associated with the proposed positive amplitude new source  \cite{wilks,mattox}.
As, in the current case, only one degree of freedom (the cross-section) is required by the DM, the TS values in 
Fig.~\ref{fig:hanetalresults} are approximately equal to the square of the number of standard deviations
of a DM detection.
 Further details on our analysis are given in the next section.

The 2FGL catalogue is fundamentally a catalogue of significant gamma-ray point sources detected by the LAT in the first 24-months of operation~\cite{2FGL}. The method employed by the Fermi Collaboration to construct the 2FGL catalogue consists of three analysis steps; source detection, localization (position refinement), and significance estimation. Transient sources were not considered. 

After applying the method outlined above, the collaboration considered a spatially unresolved gamma-ray source as part of the 2FGL catalogue if its test statistic TS exceeds 25, or equivalently (as there are four new degrees of freedom), if its significance detection was larger than $4.1\sigma$. Generally speaking, it might occur that sources that were just on the threshold for the 2-year data set, become significant for larger LAT data samples.
For example, after two years of LAT observations, Pictor A had a test statistic value of 20, however, after three years of LAT observations, Pictor A was found to have a TS value of 33~\cite{Jenni} . It is important to note that when using extended dark matter profiles, new gamma-ray sources not taken into account in the model can mimic a DM signal. Here, we show that indeed the emission from weak gamma-ray point sources, that were not included in the  2FGL catalogue,  were a significant contaminant to the flux interpreted as continuous gamma-ray emission in HFEGW.

\section{Analysis and Methods}
\label{sec:analysis}

In order to compute the effect that weak point sources (which were not included in the 2FGL catalogue) have on the analysis presented in HFEGW, we considered three different periods of Fermi data: 2 years, 3 years and 3.8 years of nominal all-sky survey data. The filters applied to the data sets, the instrument response function utilized, background models and properties of the Virgo cluster assumed are the same as those adopted in HFEGW. We refer the reader to that study for details.

To check for new point sources in the ROI we do not assume the extended DM template as part of the model. A brief description of the method used is outlined in Sec.~\ref{subsec:missingptsrc}. In Sec.~\ref{subsec:qualityfit} we illustrate the spatial and spectral agreement of the new proposed background with the observations and we then recheck for the claimed continuous DM self-annihilation in Sec.~\ref{sub:dmsignal}.   
\subsection{Determination of missing gamma-ray point sources}
\label{subsec:missingptsrc}        
Using a binned likelihood technique~\cite{mattox}, we performed an analysis of the Fermi-LAT spectrum over three different periods of time: (2008 August 4 through 2010 August 4), (2008 August 4 through 2011 August 16) and (2008 August 4 through 2012 June 26). The normalization and index for all the point sources from the 2FGL catalogue that fell within $10^{\circ}$ of the center of the core of M87 were left free in the fit. The spectral parameters of sources which were within $5^{\circ}$ of the ROI perimeter were fixed to their catalogue values.  

\begin{figure}[!t]
\begin{tabular}{cc}
              \centering
  \includegraphics[width=0.5\linewidth]{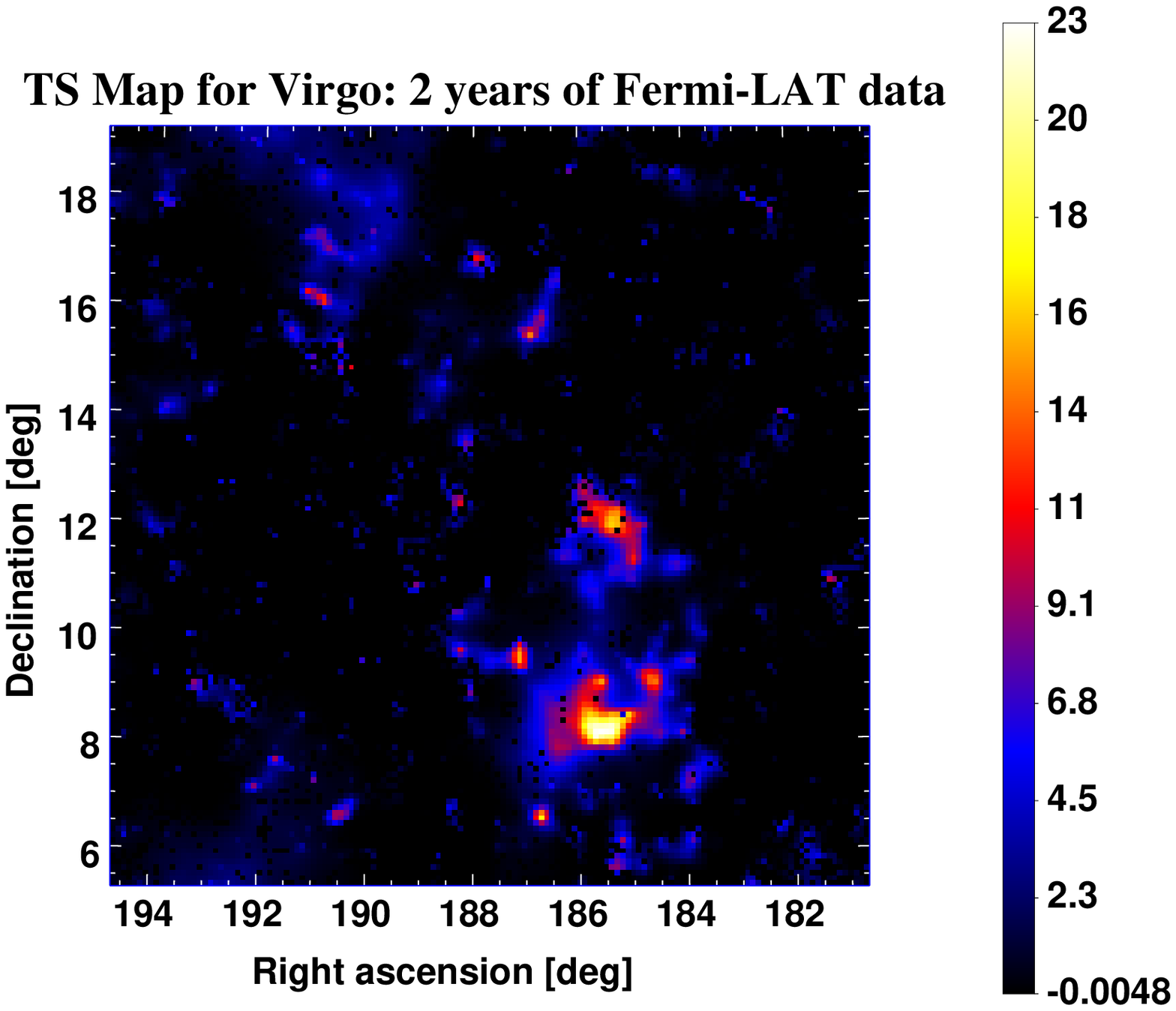} &
               \centering
  \includegraphics[width=0.5\linewidth]{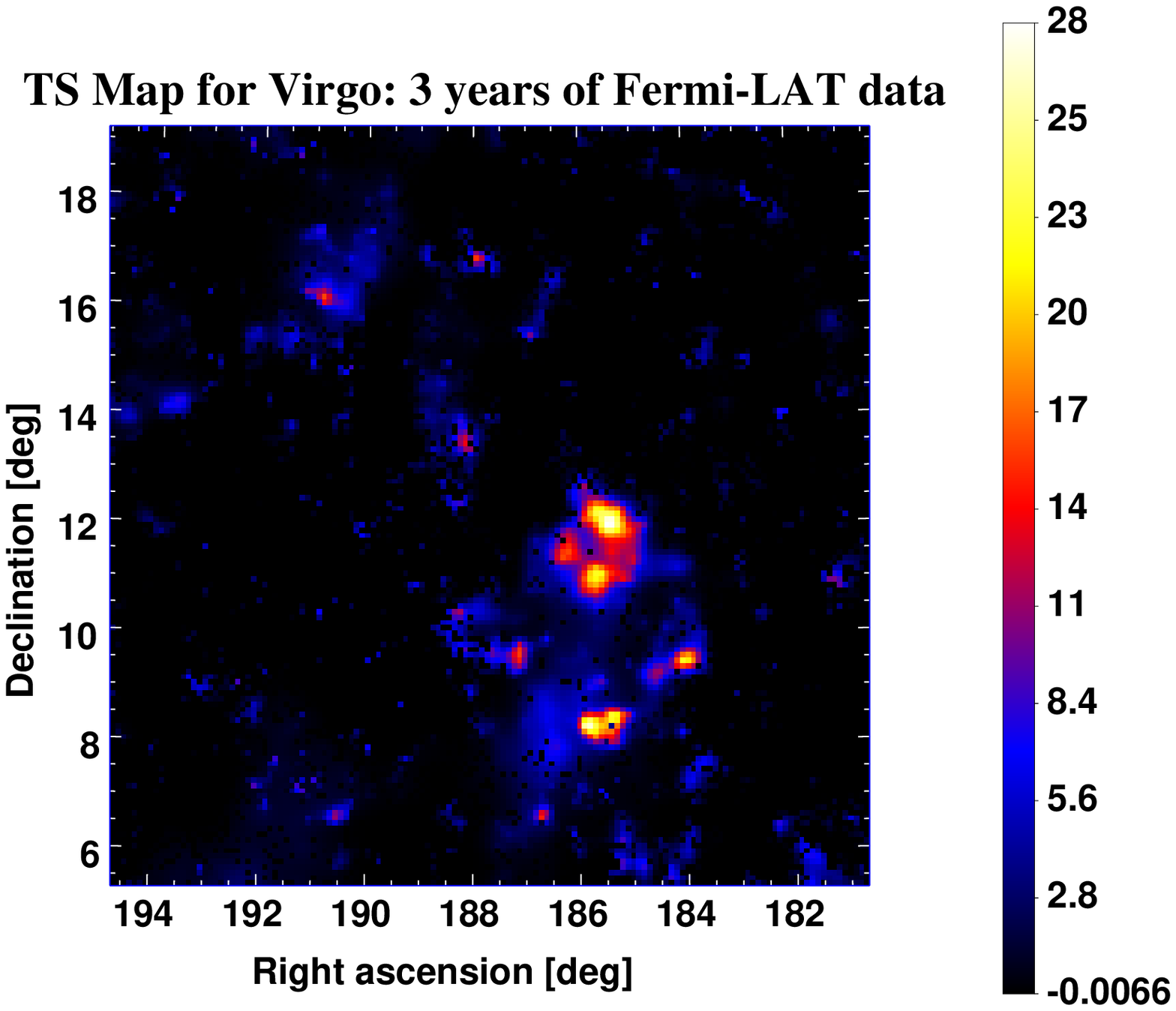} 	
\end{tabular}
\centering
  \includegraphics[width=0.5\linewidth]{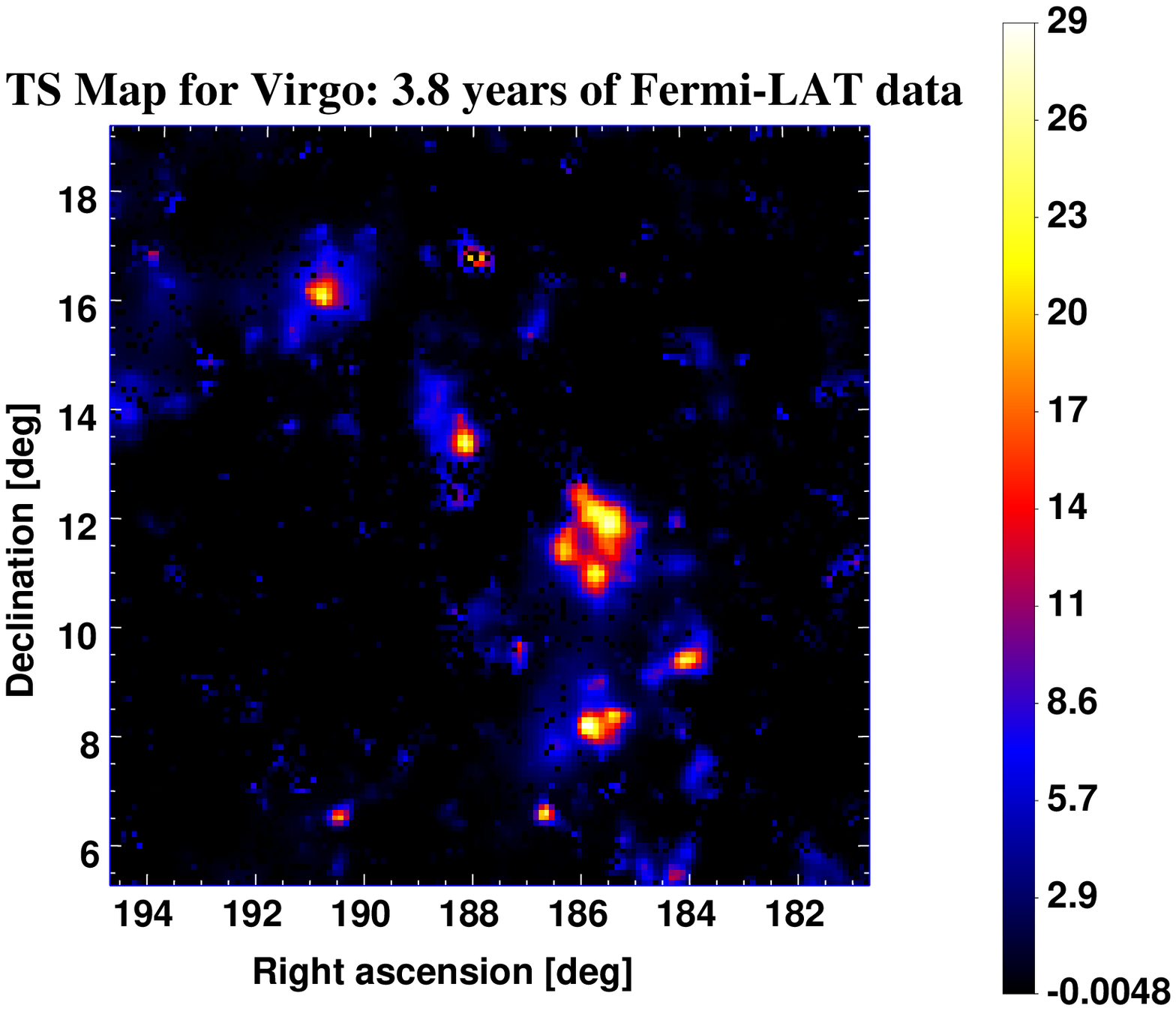}
  \caption{ \label{fig:tsmaps}TS maps for three different periods of Fermi data of the Virgo cluster. The maps span a $14^{\circ} \times 14^{\circ}$ region of the sky which is centered at the coordinates of M87. The extent of every pixel is $0.1^{\circ} \times 0.1^{\circ}$.}
\end{figure}

From the resulting best-fit we construct a residual 
TS map for each of the three data sets over  a grid of 19600 points in a $14^{\circ}\times 14^{\circ}$ squared region centered at M87. For each given point of the grid we add sequentially a new point source with a conventional spectral definition~\cite{2FGL}, and maximize the likelihood as a function of its flux. This step was realized using the Fermi Science Tool\footnote{http://fermi.gsfc.nasa.gov/ssc/data/analysis/documentation/Cicerone/} (version v9r27p1)  \texttt{gttsmap} and the resulting array of TS values is shown in Fig.~\ref{fig:tsmaps}.
To estimate the position of the new sources, we take into consideration a list of sets of adjacent pixels which satisfy the condition $TS>10$. For every set of pixels, the coordinates of its centroid were computed as an average of the pixel positions weighted by their respective TS values. All of the candidate sources which are sufficiently isolated under visual inspection are then passed to the significance and thresholding step in this iterative process. 

In order to get the significances of the possible new point sources, we assumed their spectrum was described by a simple power-law, then a binned \texttt{pyLikelihood$^1$} routine with an energy binning to 25 bins was run. We made no distinction between \textit{Front} and \textit{Back} events. The Fermi collaboration stipulated that sources with a $TS>25$ should
 be included in the catalogue of gamma-ray point sources~\cite{2FGL}. Those sources that survived this threshold analysis were then passed onto the next step which consisted of a position refinement. 
Finding the best position for the candidate sources is accomplished with the Fermi Science Tool \texttt{gtfindsrc$^1$}. By using an unbinned analysis technique, this tool seeks  the highest TS value for different positions around our initial
value based on an analysis of groups pixels.  We also estimate the uncertainties in the source positions in this step. 

We did not find evidence for new point sources with a $TS>25$ for the 2 years data set~\cite{2FGL}. However the significance for some weak source candidates was appreciably enhanced when more LAT data was included. The results are shown in Table~\ref{tab:1}. 
\begin{table}[H]
\centering
\begin{tabular}{c|c|c|c}
\hline\hline
Right Asc.\ [deg] & Dec.\ [deg] & 95\% error radius [deg] & TS  \\ \hline
190.92 & 16.21 & 0.07 & 31.47 \\
187.91& 16.88 & 0.09 & 24.90 \\
188.18& 13.56 & 0.11 & 41.92 \\
185.85 & 8.30 & 0.05 &30.92\\
186.68& 6.68 & 0.04 & 25.40 \\
185.48& 12.04 & 0.11 & 26.18 \\
184.12& 9.48 & 0.06 & 24.61 \\\hline
185.74& 11.06 & 0.09 & 18.25\\
187.15& 9.71 & 0.09 & 15.18 \\ \hline\hline
\end{tabular}
\caption{\label{tab:1}New point source candidates found in the Virgo cluster for 3.8 years of Fermi-LAT data. We show in the top part of the table seven new point sources which satisfy all the requirements to be included in upcoming catalogues. In the bottom part of the table an additional two sources that do not quite satisfy the thresholding condition are also shown.
}
\end{table}

\subsection{Quality of the background fit} 
\label{subsec:qualityfit}
Since there is sufficient evidence for a group of new gamma-ray sources in the Virgo cluster, we included these sources in our background model and evaluated the agreement with the LAT data. A new fit to the observations for the data period of 3.8 years was made where the normalizations of the galactic and isotropic backgrounds were left free, as well as, the spectral shape and normalizations for all point sources that fell within a squared region of $14^{\circ}\times 14^{\circ}$ centered at the M87 position.  

\begin{figure}[!t]
                \centering
  \includegraphics[width=.60\linewidth]{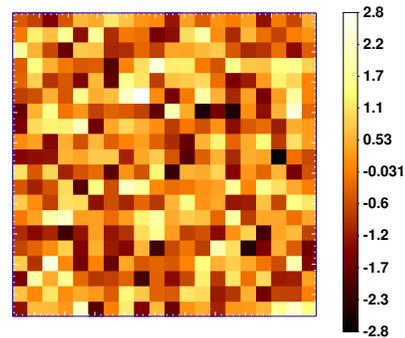}
                	\caption{ \label{fig:residuals} The spatial residual map shows $(\mbox{data}-\mbox{BG model})/\sqrt{\mbox{BG model}}$ in $\sigma$ units, where the background (BG) model includes the 2FGL catalogue sources plus the new point sources discovered in the ROI (here conservatively we consider all sources shown in Tab.~\ref{tab:1}). The pixel size was rescaled to $0.5^{\circ}\times 0.5^{\circ}$ and the map spans a $10^{\circ}\times 10^{\circ}$ centered at the cluster position. Counts are summed over the full energy range of 100 MeV$-$100 GeV for the 3.8 years of Fermi-data.}
\end{figure}

\begin{figure}[!t]

              \centering
  \includegraphics[width=1\linewidth]{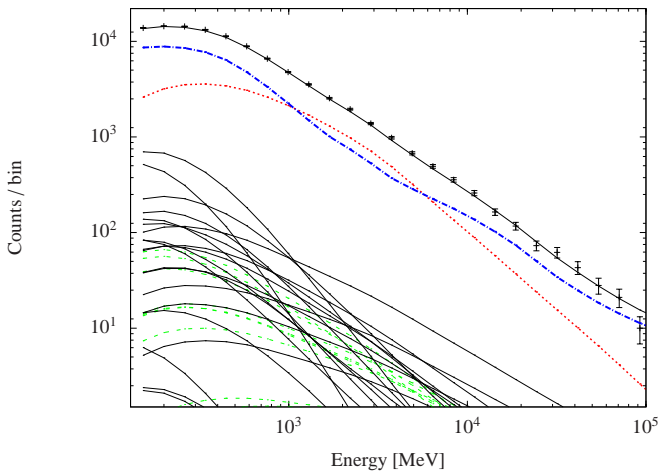} 
               \centering
\caption{ \label{fig:spectrum} Fit to the spectrum of the BG model. Red (dotted) and blue (dash-dotted) lines correspond to Galactic and isotropic extragalactic background respectively. The green dashed lines show the contribution of the new point sources and the black lines stand for the 2FGL catalogue point sources. Counts are read from a Fermi-data period of 3.8 years in the full energy range of 100 MeV$-$100 GeV.}
\end{figure}

In Figures~\ref{fig:residuals} and~\ref{fig:spectrum} we illustrate the spatial and spectral quality of the fit. The spatial residual map is scaled to the Poisson noise and the pixel size was resampled from their original size of $0.1^{\circ}\times 0.1^{\circ}$ to avoid statistical fluctuations.  It is consistent with random noise and
contains no noticeable spatial features. The spectrum in Figure ~\ref{fig:spectrum} is a good fit without any noticeable patterns in the residuals. As can be seen, the new point sources have a greater contribution to the overall fit than many of the ones already included in the 2FGL catalogue.
Thus the background model we have found is a better representation of the gamma-ray sky in the Virgo region and should therefore be used as the background template in any studies in this region~\cite{huang}. 

\subsection{The effect of unresolved point sources on the significance of the DM annihilation signal}
\label{sub:dmsignal}

The significance of a DM annihilation signal is examined using three alternative case scenarios for the background model: (i) a model following the approach taken in HFEGW for the NO-CR model; (ii) a model including the seven sources shown in the top part of Table~\ref{tab:1} with $TS \gtrsim 25$, plus the 2FGL sources and diffuse backgrounds; (iii) a model conservatively including all the new sources from Table~\ref{tab:1} plus the 2FGL sources and diffuse backgrounds.

We use the high resolution extended DM halo profile obtained in Ref.~\cite{Gao} and model the WIMP spectrum with the \texttt{DMFit} package~\cite{Jeltema} as implemented in the \texttt{Science Tools} analysis software. Since our case study is the self-annihilation of WIMP particles in the $b\bar{b}$ channel we do not take into account Inverse Compton (IC) effects. There is also the issue of whether it is possible to successfully account for the significant point source at the center of Virgo (M87).  HFEGW found that their DM signal was spatially extended and so concluded that it could not be due to the M87 point source. We also checked that M87 did not have a significantly curved or time-varying spectrum and found no vidence of extended emission
from M87 using the 3.8-year data set \cite{2FGL}.  
Based on these checks, we model M87 as a point-source with a power-law spectrum. 
Any deviation from this may erroneously enhance an apparent DM signal, but given
we find that the addition of the new point sources makes the apparent DM signal not significant, this is unlikely to be an important factor for our study.

\begin{figure}[!t]
              \centering
  \includegraphics[width=1\linewidth]{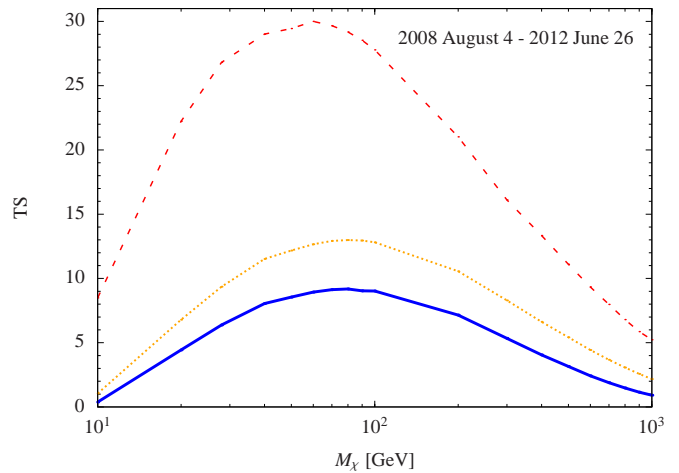} 
               \centering
\caption{ \label{fig:darkmattersignificance} TS values for DM radiation in the $b\bar{b}$ channel, an extended DM density profile and source class of Fermi-LAT data taken between 2008 August 4 and 2012 June 26. The fit is made by considering three distinct background models. The red dashed line is obtained by assuming the same background model used in HFEGW. For the yellow (dotted) and blue (solid) lines shown, the fit is obtained by using a modified background which considers 7 additional new point sources, and 9 additional point sources respectively (see text).}
\end{figure}

A new fit to the LAT data period of 3.8 years corresponding to the NO-CR model (see HFEGW for details) is shown in Fig.~\ref{fig:darkmattersignificance}. Interestingly, we note that if only the 2FGL sources plus galactic and extragalactic backgrounds are included (case (i)), the significance detection for extended DM radiation exceeds $5\sigma$. However, and the main point of this paper, all significant point sources must be included in the background model for such studies. Indeed when we included in the template model the seven new point sources with $TS \gtrsim 25$ Table~\ref{tab:1} (case(ii)) the significance of detection decreased substantially to $3.6\sigma$. And for case (iii) when we included all of the new point sources found with TS-values larger than 15 the significance of detection decreased to $3.0\sigma$. Here we should also stress that if a detailed CR component is added to the model, the significance of detection for DM would decrease further.          

As can  be seen in Fig.~\ref{fig:tsmaps}, the data favor additional localized point sources rather than a more diffuse signal that would be expected from annihilating DM. In introducing seven new point sources we should however consider that we are introducing 28 new parameters (seven times the positions, the amplitudes and the spectral indices), while adding DM corresponds to only two new parameters (the cross-section and the mass). We can statistically compare the two alternatives by evaluating the p-values for each case using Wilks' theorem \cite{wilks,mattox}. 
As the p-values are quite small we can convert to ``$\sigma$'s of detection'' by comparing the p-values to the one parameter 
case. The TS for including seven new  point sources is 192 which corresponds to a 11 $\sigma$ detection. While the TS for including DM with no new point sources corresponds to a TS of 28.9 which for two degrees of freedom is only a 5 $\sigma$ detection. So clearly the 7 new point source case is a much better fit to the data despite requiring more parameters. 

We found that the new point sources did not have significant curvature in their spectrum or time variation on a monthly scale \cite{2FGL}. There are detections of point sources at other wavelengths in areas consistent with the positions we found for the new point sources. It would be interesting in future work to evaluate statistically whether they can be associated with the new point sources, as was done for the 2FGL point sources \cite{2FGL}, but it is beyond the scope of the current article.

As the observing time increases, it is expected that more astrophysical point sources will in general be found as the signal to noise is increasing. However, it would not be valid to extrapolate the number of new point sources we have detected in the Virgo cluster to other areas of the sky. It should be noted the it was only the Virgo cluster that HFEGW found a significant signal excess despite checking several other clusters. Also, clusters in general will be expected to have a greater number of point sources compared to random areas of the sky.

\section{Conclusions}
We have investigated whether there is evidence for extended emission from DM annihilation in the Virgo cluster.
We have focussed particularly on the results found by HFEGW. There, using three years of  LAT data, they found, assuming a negligible CR contribution,  there was a 4$\sigma$ detection of a CDM component annihilating to the $b\bar{b}$ channel. They found  a somewhat weaker significance for the $\mu^+\mu^-$ case. Also, they found that including fitting for a CR component reduced the $b\bar{b}$ case  significance to  $3\sigma$. 
We have focussed on their most significant case to highlight the effects of
unresolved point sources.
Crucially, they used the point source catalogue derived from two years of  LAT data. But, we have found the extra data now means that point sources which were below the TS=25 threshold in the 2-year data, are above it in the 3-year data. Such point sources should be included in the background template for studies using the 3-year data set.

We redid the HFEGW  analysis with 3.8 (rather than 3) years of LAT data and found that the HFEGW result went up to a 5$\sigma$ detection for DM annihilating into the $b\bar{b}$ channel (with negligible CRs assumed)  if the 2FGL catalogue is still used to construct the background template. But, if we  included the seven new point sources with $TS \gtrsim 25$ for the four year data, we found the DM signal is only significant at the $3\sigma$ level, and this will go down further if the CRs are fitted for rather than assumed negligible. Therefore, we have shown that the HFEGW result
was significantly affected by unresolved point sources. This highlights that when using the LAT data, it is important to  check for new point sources if one is using more data than
was used in deriving the most recent point source catalogue.

Shortly after a preprint of the current paper was placed on the arXiv, a revised version of HFEGW was uploaded, with an extended set of authors  \cite{HFEGW2}. Their new conclusions were consistent with ours. There were some small differences in the number and positions of the new point sources found which is likely due to the slightly different algorithms used for point source detection and the area for which new sources were searched for. However, as in this paper, the conclusion was that the continuous gamma-ray signal, interpreted as evidence for DM in HFEGW, was in fact due to unresolved point sources.

\acknowledgments
O.M-R. acknowledges to Johann Cohen-Tanugi, Dave Davis, Jiaxin Han and Jeremi S. Perkins for fruitful suggestions on the software usage.
 O.M-R. specially expresses his thanks to Stefano Profumo for allowing him to study the original Fortran version of the \texttt{DMFit} package. 
 O.M-R. is supported by the UC Doctoral Scholarship.
 AMB acknowledges the financial support of the Marsden Fund Council from New Zealand Government funding, administered by the Royal Society of New Zealand.
   \\


\bibliography{references}

\end{document}